\documentclass[10pt,letterpaper]{article}
\usepackage[top=0.85in,left=1.7in,footskip=0.75in,marginparwidth=2in]{geometry}

\usepackage[utf8]{inputenc}

\usepackage{xcolor}
\usepackage{todonotes}
\usepackage{hyperref}
\usepackage[T1]{fontenc}
\usepackage{multirow}
\usepackage{cite}
\usepackage{stfloats}

\usepackage{cite}

\usepackage{nameref,hyperref}

\usepackage[right]{lineno}

\usepackage{microtype}
\DisableLigatures[f]{encoding = *, family = * }

\raggedright
\usepackage{changepage}

\usepackage[aboveskip=1pt,labelfont=bf,labelsep=period,singlelinecheck=off]{caption}

\makeatletter
\renewcommand{\@biblabel}[1]{\quad#1.}
\makeatother

\usepackage{lastpage,fancyhdr,graphicx}
\fancyheadoffset[L]{2.25in}
\fancyfootoffset[L]{2.25in}

\usepackage{color}

\definecolor{Gray}{gray}{.25}

\usepackage{graphicx}

\usepackage{sidecap}

\usepackage{wrapfig}
\usepackage[pscoord]{eso-pic}
\usepackage[fulladjust]{marginnote}
\reversemarginpar

\begin{document}
\vspace*{0.35in}

% title goes here:
\begin{flushleft}
{\Large
\textbf\newline{SciGRID\_gas - Data Model of the European Gas Transport Network.}
}
\newline
% authors go here:
\\
Adam~Pluta\textsuperscript{1},
Medjroubi~Wided\textsuperscript{2},
Jan~Diettrich\textsuperscript{1},
Jan~Dasenbrock\textsuperscript{1},
Hendrik-Pieter~Tetens\textsuperscript{2},
and~Javier~E.~Sandoval\textsuperscript{2}
\\
\bigskip
\bf{1} Institute of Networked Energy Systems\\
German Aerospace Center (DLR)\\
Oldenburg, Germany
\\
\bigskip
* Adam.Pluta@dlr.de

\end{flushleft}

\section*{Abstract}
The current transition in the European energy sector towards climate neutrality requires detailed and reliable energy system modeling. The quality and relevance of the energy system modeling highly depend on the availability and quality of model input datasets. However, detailed and reliable datasets, especially for energy infrastructure, are still missing. In this contribution, we present our approach to developing an open-source and open-data model of the gas transport network in Europe. Various freely available data sources were used to collect gas transport datasets and their attributes. The resulting datasets of the various data sources were processed, and unique elements were merged together. Statistical and heuristic methods were used to generate missing network element attributes.
\newline \indent As a result, we successfully created a gas transport network model only using open-source data. The SciGRID\_gas model contains 237.000\,km of pipeline data which is a very good approximation to know length values. In addition, datasets of compressor stations, LNG terminals, storage, production sides, gas power plants, border points, and demand time series are provided. Finally, we have discussed data gaps and how they can potentially be closed.

\section{Introduction}
The most significant shares of gross energy consumption in Europe in 2019 were held by oil and petroleum products (34.5\,\%), followed by natural gas (23.1\,\%) representing almost 60\,\% of energy consumption \cite{Eurostat-NRG_BAL_C}. The high share of natural gas in the energy mix reflects its essential role as an energy carrier and the need to decarbonize sectors where it is used as a fuel source. This will be achieved mainly by ramping up the integration of renewable energy sources (RES) in the energy system. In this context, new flexibility concepts are needed to integrate higher RES shares while maintaining energy supply stability and reliability. Such flexibilities are the Power-to-X (P2X) technologies combined with energy storage \cite{sternberg2015power}. P2X refers to various processes to convert and "store electricity" using surplus RES electric power to seasonally and spatially balance energy. A promising P2X technology is power-to-gas (P2G) which has a vast potential in decarbonizing different energy sectors such as heating and transport. P2X uses surplus electrical power to produce hydrogen via water electrolysis \cite{stockl2021optimal, kakoulaki2021, schiebahn2015power}. Hydrogen can then be used as a fuel directly or converted to LPG, syngas, or methane. The produced gas can then be transported using the current gas transport network.\\ 
Despite the importance of modeling and analysis of the gas sector and its interactions with other energy sectors, no reliable datasets describing the gas transport grid exist. Examples of available datasets are limited to single countries and do not provide details of the grid components. Such examples are the LKD-EU dataset for Germany \cite{kunz2017electricity} and the National Grid dataset for the UK \cite{GB_NationalGrid}. The lack of datasets motivated us to initiate the SciGRID\_gas project \cite{SciGRID_gas_ws} at the DLR Institute of Networked Energy Systems. The goal of the SciGRID\_gas project is to derive a reliable and detailed dataset for the gas transport grid in Europe which can be used for modeling and analysis purposes. In practice, the source code of the data model, the geo-referenced datasets describing the gas transport grid as well as the documentation, are made available under open source licences. In order to use SciGRID\_gas in the simulation of energy systems, the integration of the datasets in existing energy system models such as \textit{open\_eGo}, \textit{PyPSA}, and \textit{pandapipes} is planned.
With the SciGRID\_gas data model, we would like to answer the following research questions:
\begin{itemize}
\item Can we build a reliable data model for the European gas transport grid using only publicly available data? 
\item Is the amount of available parameter data sufficient to estimate missing parameter data via heuristics and statistical methods?
\end{itemize}

This contribution is structured as follows: In Chapter\,\ref{dataorigin}, we discuss the data sources used for constructing the open-source gas transport network model. This is followed by the discussion of the model architecture in Chapter\,\ref{model}. Chapter\,\ref{sec:methods} gives a short overview of some suitable methods for creating the model. Due to the page number limitation, more detailed information is also available in the respective model documentation, which is accessible online\cite{SciGRID_gas_ws}. 
Chapter\,\ref{sec:results} presents our model’s graphical and statistical results. This is followed by the discussion  in Chapter\,\ref{sec:discussion} and the conclusion and outlook in Chapter\,\ref{sec:conclusion}.

\section{Data sources} \label{dataorigin}

Obtaining reliable open-source data of the gas transport system is a challenging task. The grid data of gas Transmission System Operators (TSOs\footnote{The European operators of the gas transmission (transport) grid are associated in the European Network of Transmission System Operators for Gas (EntsoG)}) are commonly not standardized, nor are they freely accessible \cite{EMap_MainRef}. Data are generally not geo-referenced and mainly available as PDF maps. Most individual TSOs are not willing to share their data due to competitive reasons. Within the SciGRID\_gas project, we have gathered freely available data from different sources. 
The most relevant are presented below. In the subsection name, we first indicate the source of the dataset (ex. Web Search) followed by the name (ex. INET) we gave the dataset in the SciGRID\_gas project. 

\subsection{Web Search - INET}
 We carried out a web search on all gas network components and compiled the gathered data into the INET dataset. The data stems from TSO press releases, TSO transparency platforms, and TSO public data. Some TSO information had to be made available by the TSOs due to EU regulations \cite{EU543013}. Other information has been made public as part of a company’s self-presentation and advertisement. The collected information includes data on network components. This includes their positions but also relevant energy modeling parameters, like diameter, capacity, power, pressure, etc.   
 
\subsection{German gas model - LKD}
The \textit{long-term planning and short-term optimization} dataset (LKD)\cite{LKD_MainRef} contains geo-referenced data on gas facilities in Germany. It was created by several German research institutes and includes information on gas pipelines, production sites, storage, compressor locations, and nodes. The SciGRID\_gas project was granted the right to use, change and redistribute the LKD data under an open license from the LKD project members. 

\subsection{EntsoG - EMAP}
The development of the European gas transport network is coordinated by the European Network of Transmission System Operators for Gas (EntsoG)\footnote{https://www.entsog.eu/}. The EntsoG is an association of 44 European TSOs, three associated partners, and nine observers. EntsoG members are required to publish certain information according to EU directives. A significant amount of this information is incorporated in the freely available and regularly updated map of the gas pipelines, drilling platforms, and storage facilities. The SciGRID\_gas project extracted the rough course and location of the depicted gas pipelines, storage, and production facilities using the map\footnote{https://www.entsog.eu/maps} of 2019. 

\subsection{Eurostat - Cons}
\begin{figure}[ht]
\centering
\includegraphics[width=0.9\textwidth]{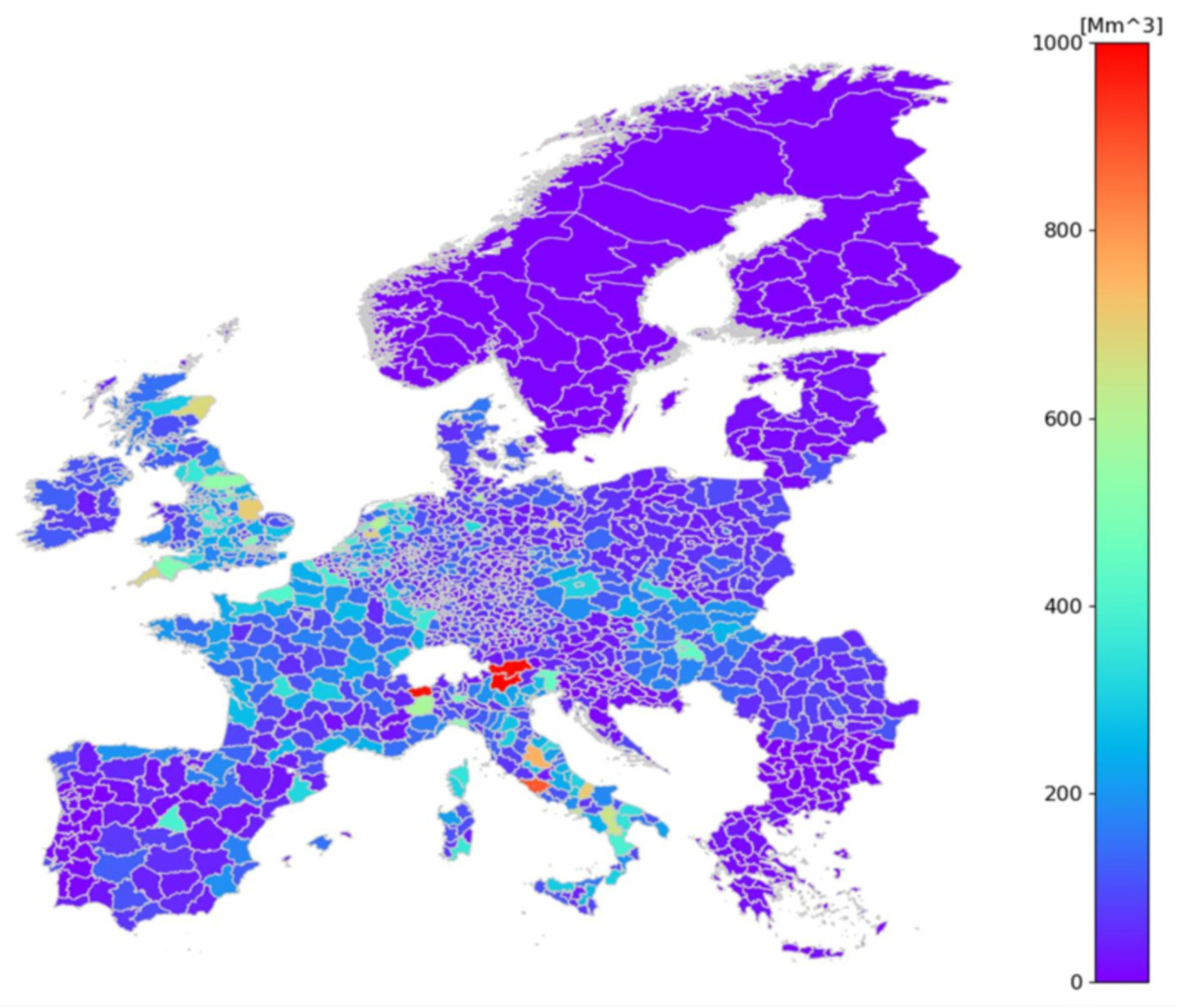}
\caption{Annually averaged gas demand in the residential sector between 2010-2019, disaggregated into Nuts 3 regions. \cite{javier}.}
\label{fig:img_mask_pred3}
\end{figure}

The European Statistical Office (Eurostat) collects and publishes data on energy supply, transformation, and consumption on a monthly and yearly basis. Eurostat statistics — like the \textit{complete energy balances} dataset \cite{EuroStat}, and others — provided the data foundation for one of our studies regarding the European gas demand. We derived gas demand time series with a daily resolution covering the years 2010-2019 with a NUTS 3 spatial resolution\footnote{NUTS is a geographical system to divide the EU territory into hierarchical levels. In Germany, NUTS 1 are federal States (Bundesländer), NUTS 2 are governmental regions known as Regierungsbezirke, and NUTS 3 are the districts (Kreise).} for 27 European countries. To provide detailed information for modelers and dataset flexibility the time series distinguishes between the sectors \textsc{households}, \textsc{commercial} and \textsc{industry}. Figure\,\ref{fig:img_mask_pred3} provides an exemplary data plot for the annually averaged residential gas demand in Europe (2010-2019) disaggregated into NUTS 3 regions. The data derivation techniques showed good benchmarking results against three existing time series of gas demand in Germany \cite{Sandoval_COMS_2021} originating from the DemandRegio project \cite{demandregio}. Significant insights were obtained when analyzing the times series concerning the seasonal, geographical, and sector-specific variability of the gas demand in Europe \cite{Sandoval_COMS_2021}.

\subsection{OpenStreetMap - OSM} 
OpenStreetMap (OSM) \cite{OSM} is a freely modifiable and accessible geo-data database with steadily increasing data coverage and data quality. In the past, OSM data has contributed to the field of energy system modeling for example in the creation of power grid models \cite{MEDJROUBI201714} or the optimization of flexibility options for urban areas \cite{Alhamwi_2017, Alhamwi_2017a}.
With the \texttt{esy-osmfilter} \cite{Pluta2020esyosmfilterA} we created a Python library to easily access and filter data from OpenStreetMap. We used \texttt{esy-osmfilter} to analyze the European gas pipeline data in OSM. Our analysis showed that data of gas transport pipeline is strongly represented and rapidly growing. However, some countries show significant data gaps. Moreover, data relating to system-relevant components like compressors stations or storage is clearly missing. With \texttt{osmscigrid}\cite{osmscigrid} we created another library to convert OSM pipeline data directly to SciGRID\_gas data format for easier integration of OSM datasets. 
OSM data has the downside of being licensed under ODbL, which is not compatible with CC-BY license, a hurdle that can be overcome using a collective database. However, we have decided that the current data will not be used directly in our model but only to validate the topology. 

\subsection{Landsat 5} 
In order to address missing data, we conducted a feasibility study on gas pipeline detection using remote sensing and Artificial Intelligence (AI) methods. We used an approach based on detecting the 16-28 m wide gas pipeline construction lane in order to detect pipeline routes. We then trained a convolutional neural network to discriminate between pixels labeled as \textsc{Pipeline} and pixels labeled as \textsc{Background}. For this purpose, we have used Landsat 5 imagery. Training and tests on the British gas transmission network and the NEL pipeline showed good evaluation scores. They proved the concept of using AI and remote sensing methods on historic open-source satellite imagery to detect pipeline pathways \cite{Dasenbrock-energies-2021}.\\

\section{Model Architecture}\label{model} 
The SciGRID\_gas data model \textbf{network} consists of several  \textbf{component classes}, each representing a list of objects. The following component classes have been implemented:
\textsc{PipeSegments} (PS),
\textsc{BorderPoints} (BP),
\textsc{Compressors} (CS),
\textsc{LNGs} (LNG),
\textsc{PowerPlants} (PP),
\textsc{Productions} (PO),
\textsc{Consumers} (CO),
\textsc{Storages} (ST). Any object which is a member of a component class is defined as an \textbf{element} of that respective class and can therefore be described by a common component-specific set of attributes. To keep track of the data processing steps, we have also included parameter information for each attribute of a specific element.

To make this data structure suitable for gas networks, we need to restructure the data  using \textbf{nodes} and \textbf{edges}. These are connected to the elements in our dataset by a unique ID. In that way, all components with the exception of \textsc{PipeSegments} are implemented as nodes. However, \textsc{PipeSegments} are also related to a start and end node, so they were implemented as edges. Intermediate pipeline points, reflecting the geographical course of \textsc{PipeSegments}, are stored in separate lists.

\section{Methodology}\label{sec:methods} 

In this section, we describe the methodology we developed to create the gas network dataset. This addresses, in particular, the release of datasets of various sources, the creation of a merged network dataset, the post-processing, and the visualization.
 
\subsection{Data Basis}
We have created and released datasets from the various data sources mentioned in Chapter$\,$\ref{dataorigin} on the project website\cite{SciGRID_gas_ws} under the section \textit{downloads}, which were converted from their original format to the SciGRID\_gas format described in Chapter$\;$\ref{model}.
Table\,\ref{tab:data origin2} gives an overview of the data sources constituting the IGGIELGNC-1 dataset.
\begin{table}[ht]
\caption{Overview of the available data sources for different gas transport components.}
\begin{center}
\begin{tabular}{l | l}
 \textbf{component} & \textbf{data source} \\
 \hline
 \textsc{PipeSegments}   & INET, EMAP, LKD, (GB, NO) \\
 \textsc{Nodes} &  INET, EMAP, LKD,  GIE, CONS \\  
 \textsc{Lngs}  &  INET, GIE \\ 
 \textsc{Storages}   & GIE, GSE, LKD, EMAP\\
 \textsc{PowerPlants}   & INET, CONS\\
 \textsc{Productions}   & INET, EMAP, LKD \\
 \textsc{Compressors}   & INET, LKD\\
 \textsc{Borderpoints}   & INET\\
 \textsc{Consumers}   & INET, CONS\\
\end{tabular}
\end{center}
\label{tab:data origin2}
\end{table}

\subsection{Data Merging Process}
In the next step, we have merged the various datasets we obtained. This task required identifying duplicate elements that exist in more than one dataset. For this task, we rely on the criteria of spatial and name similarity using the \texttt{fuzzywuzzy} python package \cite{fuzzy}. The python algorithm will evaluate the identity of objects with an identity score between 0 and 100, where 0 indicates no similarity and 100 indicates apparent duplicity. Components are merged if they exceed a component-dependent threshold between 80 and 95. In case that the likely duplicates do not share the same attributes, the attributes of the subjectively most trustworthy source from Chapter$\;$\ref{dataorigin} are adopted.

This process works for all components which are implemented as nodes. The process of merging pipelines is more complex.
For edge,s the process is built around a similarity check of the start and end node positions and comparisons of the diameter, pressure, capacity, and length values. The respective algorithm is described in more detail in the documentation of the final dataset on the website\cite{SciGRID_gas_ws}. In terms of data, we have used the pipeline data from the EMAP dataset as a basis for the pipeline system and combined them with pipeline data from the INET and LKD datasets.   

\subsection{Attribute Generation}
Once a merged dataset has been compiled, we focus on predicting missing data on the attribute level. Depending on the specific attribute, various approaches produced different results regarding their suitability to estimate missing values. 
One approach would be to use heuristics to determine missing values. For example, in the case of missing pipeline capacity values, one could use the capacity of an adjacent compressor station to derive this value under the consideration of all other incoming and outgoing pipelines. This approach, of course, requires the construction of meaningful heuristics and sufficient data.

Another approach is exploiting linear relations between different parameters of the same element. For example, one can identify a linear regression between maximal power and maximal capacity of a compressor. For this purpose, we have used the Lasso-linear regression method from \texttt{scikit}\cite{scikit_Reg}.
However, a meaningful linear correlation is not identifiable in most cases, or the data density is insufficient. Thus, we must rely on a statistical approach and calculate mean or median values. 

We have also used some of our unused data from Chapter$\;$\ref{dataorigin} to derive statistical correlations and heuristics in some particular situations. For all derived values, the data generation process will also store the applied method in the corresponding metadata of this value. This is especially useful if a user wants to distinguish raw and generated data. The user can identify  further heuristics and develop other data generation methods based only on the original attribute data.

\subsection{Post-Processing and Visualization}

Finally, we have added our artificial \textsc{Consumer} (or \textsc{demand}) nodes to the network and connected them to the nearest pipeline. We have compiled the final dataset for the three different consumer aggregation levels NUTS 1, NUTS 2, and NUTS 3, which resulted in the datasets: IGGIELGNC-1 \cite{IGGIELGNC1}, IGGIELGNC-2 \cite{IGGIELGNC2} and IGGIELGNC-3 \cite{IGGIELGNC}; respectively. Further, some cleanup routines have been implemented, e.g., for removing or connecting unconnected isolated elements to create a coherent network. Also, during post-processing, the elevation of each node was determined with the help of \textit{Bing Maps Elevation API}\cite{BingAPI}. Additionally, we have released \texttt{qplot}\cite{qplot}, a \texttt{matplotlib} \cite{Hunter:2007} based visualization library for SciGRID\_gas data. The library was used for the creation of Fig.\,\ref{fig:ES-OSM}. 

\section{Results} \label{sec:results} 

We have released our final results under the name IGGIELGNC-1 on our website\cite{SciGRID_gas_ws}, where it is linked to its respective \textsc{Zenodo} repository. The data is licensed under CC-BY and available in CSV and geojson formats, and  accompanied by a methodical documentation. We want to emphasize that the following results are from version 1.1 of the dataset and are subject to changes in future updates. 

\begin{figure}[ht]
\centering
\includegraphics[width=1\textwidth]{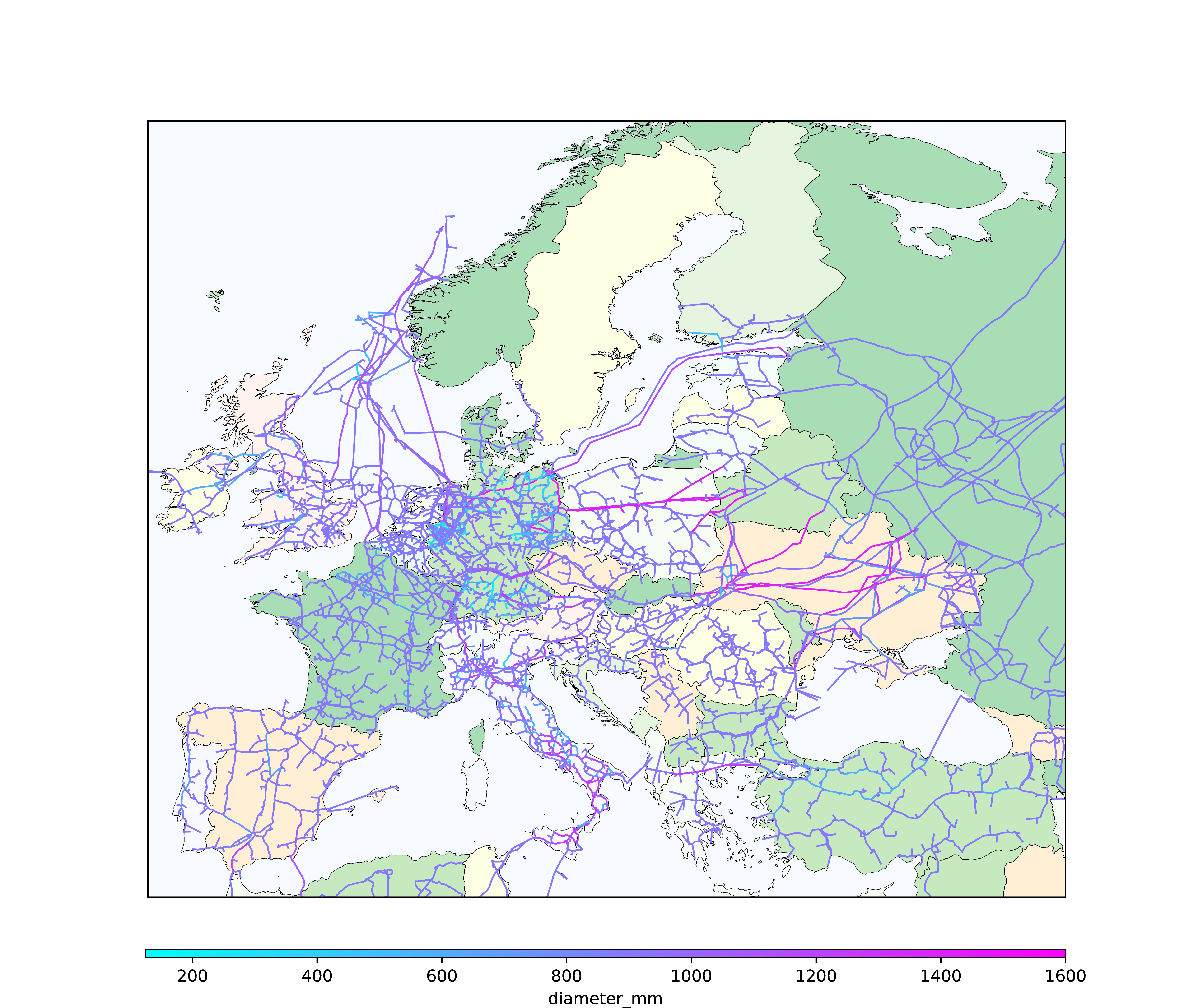}
\caption{The pipeline system of the European gas transport grid in the IGGIELGNC-1 dataset. Pipelines are colored according to their diameter values.}
\label{fig:IGG3_1}
\end{figure}

\begin{figure}[ht]
\centering
\includegraphics[width=1\textwidth]{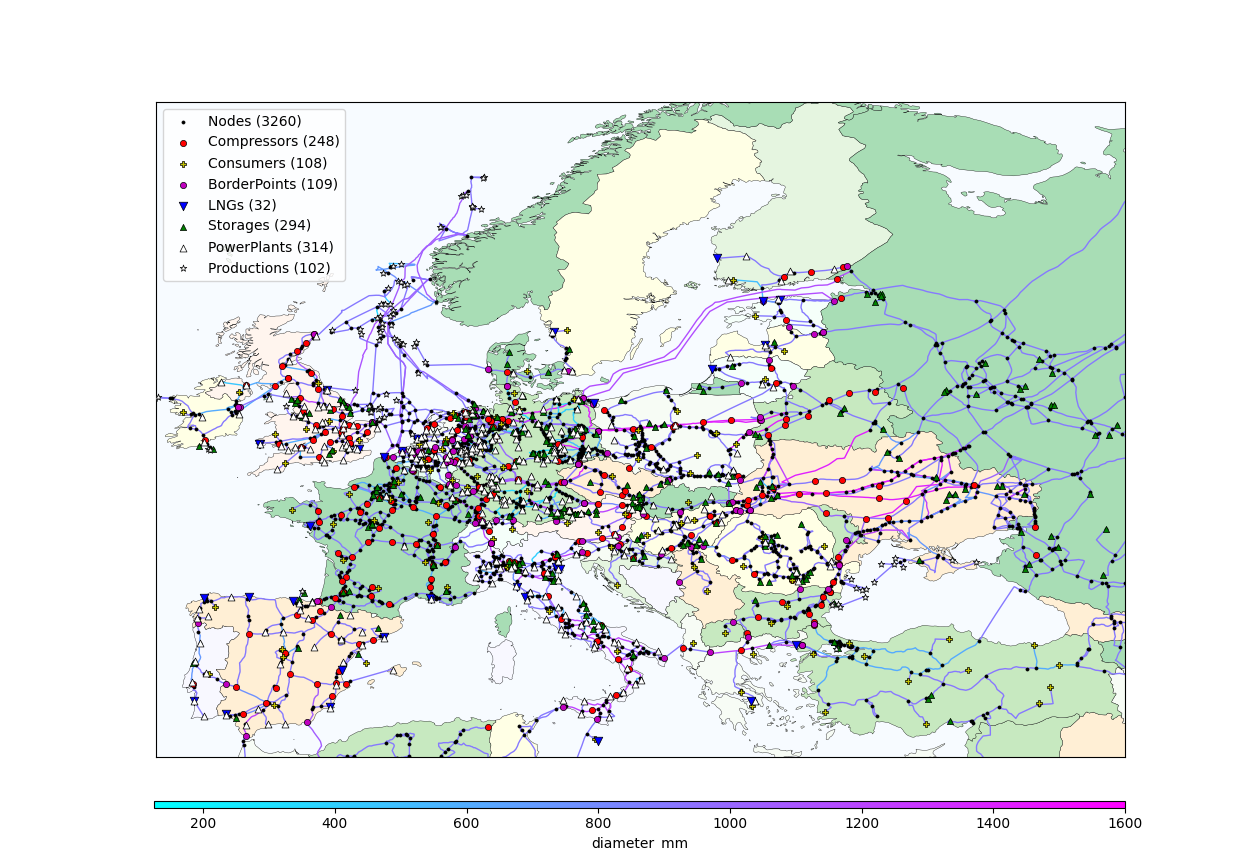}
\caption{Extract of the European gas network of the IGGIELGNC-1 dataset with all its components.}
\label{fig:IGG3_2}
\end{figure}

We have merged the pipeline systems of INET with a total length of about 60.000\,km, EMAP with a total length of about 207.000\,km, and LKD with a total length of about 27.000\,km to a network which finally contains 237.000\,km of gas transport pipelines. This pipeline system is plotted in Fig.\,\ref{fig:IGG3_1}. For comparison, the extrapolated pipeline validation dataset from OSM in 2020 only contains a total length of 108.000\,km.
In Fig.\,\ref{fig:IGG3_2} we show the pipeline system with all other components that sum up to 109 BP, 248 CS, 32 LNG, 314 PP, 102 PD, 108 CO, and 294 ST (refer to Section\,\ref{model} for the nomenclature). A country-wise overview of all components for some EU countries is shown in Table\,\ref{tab:pipe country} .

\begin{table}[htbp]
\caption{Total pipeline (PS) length and compressor station (CS) count for some countries. Data source: IGGIELGNC-1 }
\begin{center}
\begin{tabular}{c | c | c |c | c | c | c}
country & PS & CS & LNG & BP & ST & CO\\
 code   & length & count &count&count&count&count \\
\hline
AT &  2451 &   7 &   0 &   4 &  15 &   3 \\
BE &  2312 &   6 &   1 &   6 &   1 &   3 \\
CH &  1012 &   1 &   0 &   2 &   0 &   1 \\
CZ &  2159 &   6 &   0 &   3 &  10 &   1 \\
DE & 27708 &  35 &   0 &  15 &  68 &  15 \\
DK &   841 &   1 &   0 &   1 &   3 &   1 \\
ES &  8389 &  18 &   7 &   4 &   8 &   5 \\
FR & 15424 &  40 &   4 &   7 &  23 &  12 \\
GB &  6836 &  28 &   4 &   4 &  22 &  11 \\
IT & 12053 &  14 &   4 &   6 &  20 &   4 \\
\end{tabular}
\end{center}
\label{tab:pipe country}
\end{table}

Next, we looked at the attribute data for \textsc{PipeSegments}, \textsc{Storages}, \textsc{LNGs} and \textsc{Compressors}. For this purpose, we have chosen up to three of the most relevant attributes for each component and determined their respective data density. Tab.\,\ref{tab:parameter} shows the result in percent. We have not considered \textsc{BorderPoints} or \textsc{Consumers} for this analysis as these components are based on data aggregation and have no real physical counterpart.

\begin{table}[ht]
\caption{Raw parameter density of the IGGIELGNC-1 dataset.}
\begin{center}
\begin{tabular}{l | c | c| c}
 \textbf{component} & \textbf{density} & \textbf{density} & \textbf{density} \\
 \hline
 \multirow{ 2}{*}{\textsc{PipeSegments}}& capacity &diameter& pressure \\
 & 13\% & 32\% & 19\% \\
 \hline
  \multirow{ 2}{*}{ \textsc{Lngs}} & capacity &size& \multirow{ 2}{*}{-}\\
 & 94\% & 69\% &  \\
 \hline
  \multirow{ 2}{*}{ \textsc{Storages}}  &capacity& power &  pressure\\
 & 63\% & 28\% & 35\% \\
 \hline
   \multirow{ 2}{*}{\textsc{Compressors}}  &capacity & power & pressure\\
 & 7\% & 15\% & 7\% \\
  \hline
    \multirow{ 2}{*}{\textsc{PowerPlants}}  &energy &\multirow{ 2}{*}{-}  & \multirow{ 2}{*}{-}\\
 & 100\% &  &  \\
  \hline
    \multirow{ 2}{*}{\textsc{Productions}}  &supply &\multirow{ 2}{*}{-}   &\multirow{ 2}{*}{-} \\
 & 5\% & &   \\
\end{tabular}
\end{center}
\label{tab:parameter}
\end{table}

\begin{figure}[ht]
\centering
\includegraphics[width=1\textwidth]{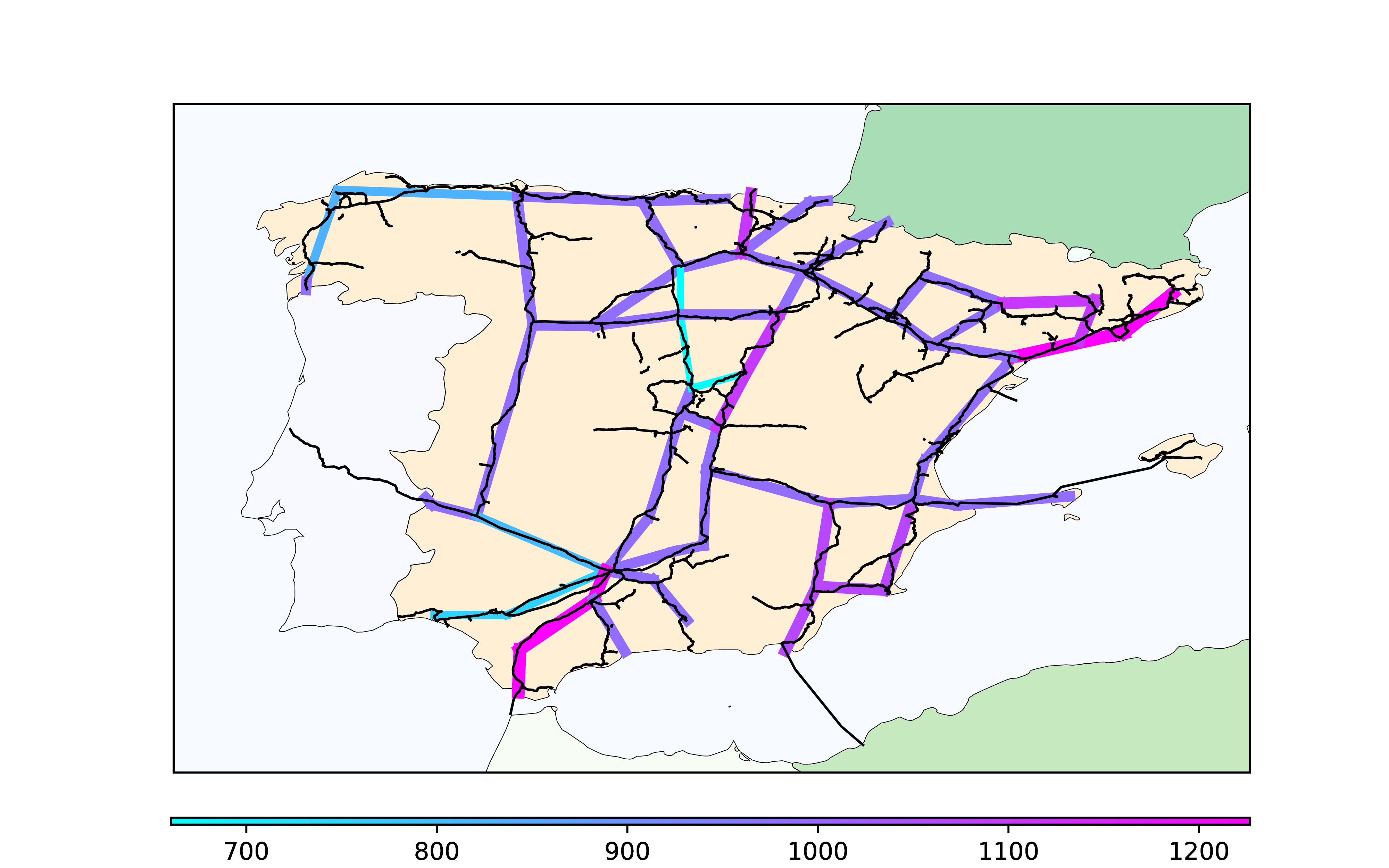}
\caption{Comparison of pipelines from OSM (black) and the INET dataset (colored with diameter [mm]).}
\label{fig:ES-OSM}
\end{figure}

We have visually validated the topology of our network datasets with the OSM pipeline data. This process is illustrated in Fig.\,\ref{fig:ES-OSM} for the validation of the INET data for the region of Spain. Our overall impression was that the topology of all major pipelines is in good agreement with the currently available OSM data.

\section{Discussion} \label{sec:discussion}

We presented our approach to create a gas transport network data model of Europe from publicly available data sources. In terms of pipelines, our data model has a total length of about 237.000\,km, which is roughly in accordance with the commonly assumed length of 200.000\,km \cite{CONNOLLY2014475}. The slight overestimation is probably a direct result of our broader definition of the European gas network, which also incorporates the Western part of Russia and some Northern African States and Turkey. However, the total length is a good indicator for the overall success of modeling the European gas transport network from open-source data. 

Further, we have used OpenStreetMap data to validate our network's topology visually. Our data show good accordance in this regard. Nevertheless, the OSM dataset currently contains only about 45\,\% of all pipelines due to significant data gaps in some regions. Also, this method can not be used to validate parallel pipelines because they are not explicitly specified in OSM. The examination of other components like \textsc{Powerplants} and \textsc{Productions} has shown data gaps in some regions stemming from incomplete data sources. We have further noticed that some countries still have missing \textsc{Borderpoints}. Since our algorithm generates these artificial nodes, this will be updated in future dataset versions.

Further validation of our dataset is currently not feasible due to a lack of relevant open-source validation datasets. This is also why we have avoided evaluating our attribute generation methods, which are described in more detail in the dataset documentation. We rather want to discuss the potential accuracy of such methods. The results of any attribute generation method, designed to predict partially unknown data, will scale in accuracy with the percentage of known data. We have analyzed our data regarding important parameters of different network components. We therefore can state that the generated attributes for \textsc{LNGs}, \textsc{Storages} and \textsc{PowerPlants} are more trustworthy, than for \textsc{PipeSegments}, \textsc{Compressors} and \textsc{Productions}. 

From our perspective, more focus needs to be put into the data acquisition for these components. Such data might be provided by OpenStreetMap (OSM) soon. Pluta and Lünsdorf\cite{Pluta2020esyosmfilterA} have stated that the gas transport data content was rapidly growing between 2014-2019. If this trend continues or will even be supported by TSOs, OSM might become a good source for this data. At some point, it might even be possible to create an entire network from OSM data like it was done for the open-source power transmission dataset SciGRID\_power \cite{SciGRID_power_ws}. The reason why this is currently not possible for the gas grid is mainly rooted in the fact that gas transport pipelines are buried underground. This makes the direct identification of their position and additional properties difficult for OpenStreetMap mappers. 

\section{Conclusion and Outlook} \label{sec:conclusion}
This contribution shows that creating a European gas transport network model is possible using only open-source data sources. Further, we have used statistical and heuristic methods to generate missing network elements attributes. Nevertheless, complete data validation was not possible due to the lack of verification data. However, our analysis of the underlying component attribute data showed that gas transport pipelines and compressor station data show low density in terms of their main attributes. Since both components are critical in modeling gas flows, future work will diminish these data gaps. We have discussed how some data gaps can potentially be closed either by a steady growth of OSM data or by remote sensing methods. However, some data, especially on the pipeline materials and roughness values, are not accessible without the assistance of TSOs. 
We believe that our data model is a valid approximation of the current European gas network. It can potentially encourage TSOs to make their data open-source, which in the long term will result in a more precise representation of the gas grid and more suited energy scenarios. 

\section*{Acknowledgment}
The authors like to acknowledge the contribution of Ontje Lünsdorf and Alaa Alhamwi.

\section*{Funding}

This research was funded under the “SciGRID\_gas” project by the German Federal Ministry for Economic Affairs and Energy (BMWi) within the funding of the 6. Energieforschungsprogramm der Bundesregierung. Funding Code: 03ET4063.

\bibliographystyle{alpha}
\bibliography{bibliography}

\newcommand{\etalchar}[1]{$^{#1}$}
\begin{thebibliography}{AMVA17b}

\bibitem[AMVA17a]{Alhamwi_2017}
Alaa Alhamwi, Wided Medjroubi, Thomas Vogt, and Carsten Agert.
\newblock {GIS}-based urban energy systems models and tools: Introducing a
  model for the optimisation of flexibilisation technologies in urban areas.
\newblock {\em Applied Energy}, 191:1--9, apr 2017.

\bibitem[AMVA17b]{Alhamwi_2017a}
Alaa Alhamwi, Wided Medjroubi, Thomas Vogt, and Carsten Agert.
\newblock Openstreetmap data in modelling the urban energy infrastructure: a
  first assessment and analysis.
\newblock In {\em Proceedings of the 9th International Conference on Applied
  Energy}, volume 142, page 1968–1976. Elsevier, 2017.

\bibitem[AO20]{osmscigrid}
A.Pluta and O.Lünsdorf.
\newblock {\em {geopy}: Python library to create a SciGRID\_gas Pipeline
  dataset from OpenStreetMap}, 2020.

\bibitem[CLM{\etalchar{+}}14]{CONNOLLY2014475}
D.~Connolly, H.~Lund, B.V. Mathiesen, S.~Werner, B.~Möller, U.~Persson,
  T.~Boermans, D.~Trier, P.A. Østergaard, and S.~Nielsen.
\newblock Heat roadmap europe: Combining district heating with heat savings to
  decarbonise the eu energy system.
\newblock {\em Energy Policy}, 65:475--489, 2014.

\bibitem[DPS{\etalchar{+}}21a]{IGGIELGNC1}
J.C. Diettrich, A.~Pluta, J.E. Sandoval, J.~Dasenbrock, and W.~Medjroubi.
\newblock Scigrid\_gas: The final iggielgnc-1 gas transmission network data
  set.
\newblock \url{https://doi.org/10.5281/zenodo.5017621}, July, 2021.

\bibitem[DPS{\etalchar{+}}21b]{IGGIELGNC2}
J.C. Diettrich, A.~Pluta, J.E. Sandoval, J.~Dasenbrock, and W.~Medjroubi.
\newblock Scigrid\_gas: The final iggielgnc-2 gas transmission network data
  set.
\newblock \url{https://doi.org/10.5281/zenodo.5017641}, July, 2021.

\bibitem[DPS{\etalchar{+}}21c]{IGGIELGNC}
J.C. Diettrich, A.~Pluta, J.E. Sandoval, J.~Dasenbrock, and W.~Medjroubi.
\newblock Scigrid\_gas: The final iggielgnc-3 gas transmission network data
  set.
\newblock \url{https://doi.org/10.5281/zenodo.4922529}, July, 2021.

\bibitem[DPZM21]{Dasenbrock-energies-2021}
Jan Dasenbrock, Adam Pluta, Matthias Zech, and Wided Medjroubi.
\newblock Detecting pipeline pathways in landsat 5 satellite images with deep
  learning.
\newblock {\em Energies}, 14(18), 2021.

\bibitem[{Ent}20]{EMap_MainRef}
{EntsoG}.
\newblock {Home page of EntsoG}.
\newblock \url{https://www.entsog.eu/}, 2020.
\newblock Accessed: 2020-03-03.

\bibitem[Eur21a]{EuroStat}
EuroStat.
\newblock Complete energy balances.
\newblock
  \url{https://appsso.eurostat.ec.europa.eu/nui/show.do?dataset=nrg_bal_c&lang=en},
  2021.
\newblock Accessed: 2021-02-08.

\bibitem[{Eur}21b]{Eurostat-NRG_BAL_C}
{Eurostat}.
\newblock {Complete energy balances: Gross inland energy consumption}.
\newblock
  \url{https://ec.europa.eu/eurostat/databrowser/view/nrg_bal_c/default/table?lang=en},
  2021.
\newblock Online data code: NRG\_BAL\_C. Last update: 06/06/2021-23:00.
  Accessed: 2020-01-01.

\bibitem[{F. }20]{demandregio}
{F. Gotzens, B. Gillessen, S. Burges, W. Hennings, J. Müller-Kirchenbauer, S.
  Seim, P. Verwiebe, T. Schmid, F. Jetter and T. Limmer}.
\newblock {DemandRegio: Harmonisierung und Entwicklung von Verfahren zur
  regionalen und zeitlichen Auflösung von Energienachfragen}, 2020.

\bibitem[Hel18]{OSM}
D.~Helle.
\newblock {OpenStreetMap - Deutschland}.
\newblock \url{https://www.openstreetmap.de/}, 2018.
\newblock Accessed: 2019-12-12.

\bibitem[Hun07]{Hunter:2007}
J.~D. Hunter.
\newblock Matplotlib: A 2d graphics environment.
\newblock {\em Computing in Science \& Engineering}, 9(3):90--95, 2007.

\bibitem[Inc14]{fuzzy}
SeatGeek Inc.
\newblock {\em {fuzzywuzzy}: Fuzzy String Matching in Python}, 2014.

\bibitem[KKS{\etalchar{+}}17a]{LKD_MainRef}
F.~Kunz, M.~Kendziorski, W.-P. Schill, J.~Weibezahn, J.~Zepter, C.~von
  Hirschhausen, and P.~Hauser.
\newblock {\em {Electricity, Heat, and Gas Sector Data for Modeling the German
  System}}.
\newblock Deutsches Institut für Wirtschaftsforschung, Daten Dokumentation 92,
  Berlin, 2017.

\bibitem[KKS{\etalchar{+}}17b]{kunz2017electricity}
Friedrich Kunz, Mario Kendziorski, Wolf-Peter Schill, Jens Weibezahn, Jan
  Zepter, Christian~R von Hirschhausen, Philipp Hauser, Matthias Zech, Dominik
  M{\"o}st, Sina Heidari, et~al.
\newblock Electricity, heat, and gas sector data for modeling the german
  system.
\newblock Technical report, DIW Data Documentation, 2017.

\bibitem[KKT{\etalchar{+}}21]{kakoulaki2021}
Georgia Kakoulaki, Ioannis Kougias, Nigel Taylor, Francesco Dolci, J~Moya, and
  Arnulf J{\"a}ger-Waldau.
\newblock Green hydrogen in europe--a regional assessment: Substituting
  existing production with electrolysis powered by renewables.
\newblock {\em Energy Conversion and Management}, 228:113649, 2021.

\bibitem[Mic20]{BingAPI}
Microsoft.
\newblock Bing maps api.
\newblock \url{https://www.microsoft.com/en-us/maps/licensing}, 2020.
\newblock Accessed: 2018 to 2021.

\bibitem[MMK16]{SciGRID_power_ws}
C.~Matke, W.~Medjroubi, and D.~Kleinhans.
\newblock {SciGRID} - {A}n {O}pen {S}ource {R}eference {M}odel for the
  {E}uropean {T}ransmission {N}etwork (v0.2).
\newblock \url{https://power.scigrid.de}, 2016.
\newblock Accessed: 2019-09-09.

\bibitem[MMS{\etalchar{+}}17]{MEDJROUBI201714}
Wided Medjroubi, Ulf~Philipp Müller, Malte Scharf, Carsten Matke, and David
  Kleinhans.
\newblock Open data in power grid modelling: New approaches towards transparent
  grid models.
\newblock {\em Energy Reports}, 3:14 -- 21, 2017.

\bibitem[{nat}20]{GB_NationalGrid}
{nationalGrid}.
\newblock {Home page of National Grid UK}.
\newblock \url{https://www.nationalgrid.com/uk/}, 2020.
\newblock Accessed: 2018-10-01.

\bibitem[PL20]{Pluta2020esyosmfilterA}
A.~Pluta and Ontje L{\"u}nsdorf.
\newblock esy-osmfilter: A python library to efficiently extract openstreetmap
  data.
\newblock {\em Journal of open research software}, 8, 2020.

\bibitem[Plu20]{qplot}
A.~Pluta.
\newblock qplot - {A} {S}cigrid\_gas {V}isualisation {L}ibrary.
\newblock \url{https://gitlab.com/dlr-ve-esy/qplot}, 2020.

\bibitem[PM18]{SciGRID_gas_ws}
A.~Pluta and W.~Medjroubi.
\newblock {SciGRID\_gas} - open source reference model of european gas
  transport networks for scientific studies on sector coupling.
\newblock \url{https://gas.scigrid.de}, 2018.
\newblock Accessed: 2019-09-09.

\bibitem[San21a]{Sandoval_COMS_2021}
Javier~Enrique Sandoval.
\newblock Estimation and simulation of a gas demand time series for the
  european nuts 3 regions.
\newblock Master's thesis, Carl von Ossietzky Universität Oldenburg, Germany,
  Fak. 5, Institute of Physics (PPRE) D-26111 Oldenburg, Germany, 6 2021.
\newblock Supervised by Prof. Dr. Carsten Agert, Dr. Herena Torio and Dr. Wided
  Medjroubi.

\bibitem[San21b]{javier}
Javier~Enrique Sandoval.
\newblock Estimation and simulation of a gas demand time series for the
  european nuts 3 regions.
\newblock Master's thesis, Carl von Ossietzky Universität Oldenburg, 2021.

\bibitem[SB15]{sternberg2015power}
Andr{\'e} Sternberg and Andr{\'e} Bardow.
\newblock Power-to-what?--environmental assessment of energy storage systems.
\newblock {\em Energy \& Environmental Science}, 8(2):389--400, 2015.

\bibitem[SGR{\etalchar{+}}15]{schiebahn2015power}
Sebastian Schiebahn, Thomas Grube, Martin Robinius, Vanessa Tietze, Bhunesh
  Kumar, and Detlef Stolten.
\newblock Power to gas: Technological overview, systems analysis and economic
  assessment for a case study in germany.
\newblock {\em International journal of hydrogen energy}, 40(12):4285--4294,
  2015.

\bibitem[sl19]{scikit_Reg}
scikit learn.
\newblock {1.1. Linear Models (scikit learn)}.
\newblock \url{https://scikit-learn.org/stable/modules/linear_model.html},
  2019.
\newblock Accessed: 2019-08-08.

\bibitem[SSZ21]{stockl2021optimal}
Fabian St{\"o}ckl, Wolf-Peter Schill, and Alexander Zerrahn.
\newblock Optimal supply chains and power sector benefits of green hydrogen.
\newblock {\em Scientific Reports}, 11(1):1--14, 2021.

\bibitem[{The}13]{EU543013}
{The European Commission}.
\newblock {Commission Regulation (EU) No 543/2013 of 14 June 2013 on submission
  and publication of data in electricity markets and amending Annex I to
  Regulation (EC) No 714/2009 of the European Parliament and of the Council}.
\newblock
  \url{https://eur-lex.europa.eu/LexUriServ/LexUriServ.do?uri=OJ:L:2013:163:0001:0012:EN:PDF},
  2013.

\end{thebibliography}

\end{document}